\documentclass[aps,12pt]{revtex4}

\usepackage{graphicx}
\usepackage{subfigure}
\usepackage{amsmath}
\usepackage{bm}

\begin{document}

\title{Kinetic theory for radiation interacting with sound waves in ultrarelativistic pair plasmas}

\author{Mattias Marklund}
\email{mattias.marklund@physics.umu.se}
\altaffiliation[Also at: ]{Centre for Fundamental Physics, Rutherford Appleton Laboratory,
  Chilton, Didcot, Oxon OX11 OQX, U.K.}
\affiliation{Centre for Nonlinear Physics, Department of Physics, 
Ume\aa\ University, SE--901 87 Ume\aa, Sweden}

\author{Padma K. Shukla}
\altaffiliation{Institut f\"ur Theoretische Physik IV and Centre for Plasma Science 
  and Astrophysics, Fakult\"at f\"ur Physik und Astronomie, Ruhr-Universit\"at Bochum, 
  D--44780 Bochum, Germany}
\altaffiliation[Also at: ]{Centre for Fundamental Physics, Rutherford Appleton Laboratory,
  Chilton, Didcot, Oxon OX11 OQX, U.K.}
\affiliation{Centre for Nonlinear Physics, Department of Physics, 
Ume\aa\ University, SE--901 87 Ume\aa, Sweden}

\author{Lennart Stenflo}
\affiliation{Centre for Nonlinear Physics, Department of Physics, 
Ume\aa\ University, SE--901 87 Ume\aa, Sweden}

\begin{abstract}
A kinetic theory for radiation interacting with sound waves in 
an ultrarelativistic electron--positron plasma is developed. 
It is shown that the effect of a spatial spectral broadening of the electromagnetic
pulse is to introduce a reduction of the growth rates for the decay and modulational
instabilities. Such spectral broadening could be due to a finite pulse coherence length, or
through the use of random phase filters, and would stabilize the propagation
of electromagnetic pulses.  
\end{abstract}

\maketitle

The physics of electron--positron plasmas is important for understanding the  
pulsar environments \cite{asseo,beskin-etal} and laboratory plasmas irradiated by intense lasers 
\cite{berezhiani,mourou-etal}, where quantum field effects come into play \cite{marklund-shukla}.
Pair plasmas are also believed to be important in the early universe, in supernova remnants and
active galactic nuclei, and in gamma-ray bursts \cite{Piran,bengt}. A pair plasma can be created in these 
environments by collisions between strongly accelerated particles \cite{bingham}. There is also a possibility of 
pair creation via high-energy curvature radiation photons in the vicinity of strongly magnetized 
astrophysical objects, triggered by charged particles streaming along the curved magnetic field \cite{Sturrock}. 
This results in large quantities of positrons produced close to pulsar polar caps \cite{Arons,Michel}.
In laboratory environments, laser experiments with focal spot intensities exceeding $10^{20}\,\mathrm{W/cm}^2$ 
have demonstrated the production of MeV electrons and evidence of positrons via laser multiphoton--gamma 
photon interactions \cite{burke-etal,bamber-etal} as well as via electron collisions \cite{Campbell,Cowan}.
 
When the intensity of the radiation is increased, the pair plasma particles attain relativistic
quiver velocities, and the plasma dynamics turns out to be essentially nonlinear, 
see e.g.\ Refs.\ \cite{stenflo,shukla-etal,popel,shukla}.  New decay instabilities, which are 
due to relativistic effects, can thus appear \cite{stenflo1}. 
The nonlinear propagation of radiation in electron--positron plasmas has therefore been suggested to be 
the reason behind the high effective temperatures of pulsar radio emissions \cite{lominadze-etal} 
as well as a possible source of the large frequency shifts in such emissions \cite{gedalin-etal}. 

In the present paper, we will study the effects of partial coherence on the nonlinear radiation propagation 
in pair plasmas. It is shown that a spatial spectral broadening of the electromagnetic pulse can lead to 
a reduced growth rate for the decay and modulational instabilities. This type of pulse broadening could be 
due to a finite pulse coherence length (as in ultra-short laser pulse application), or through the use of 
random phase filters. The effect of the broadening is to stabilize the propagation of electromagnetic pulses 
in pair plasmas.  

The interaction between electromagnetic radiation and sound waves in pair plasmas with 
ultrarelativistic temperatures is given by the system of equations \cite{stenflo-shukla}

\begin{equation}\label{eq:radiation}
  \left( \partial_t^2 - c^2\nabla^2 + 2\omega_{\mathrm{p}}^2 \right)\bm{A} 
    + \omega_{\rm p}^2\left(\frac{4}{3}\frac{n_1}{n_0} 
      - \frac{e^2}{m^2(T)c^4} |\bm{A}|^2\right)\bm{A} = 0 ,
\end{equation}
and

\begin{equation}\label{eq:sound}
  \left( \partial_t^2 - u^2\nabla^2 \right)\frac{n_1}{n_0} = \frac{e^2c^2}{32T^2}\nabla^2|\bm{A}|^2 ,
\end{equation}
where $\bm{A}$ is the vector potential of the radiation, $\omega_{\rm p} =  (\pi n_0e^2c^2/T)^{1/2}$ is the effective 
plasma frequency, $m(T) = 4T/c^2$ is the effective mass, $T$ is the electron and positron temperature, $n_0$ is 
the unperturbed plasma density, $n_1$ ($\ll n_0$) is the density perturbation, $c$ is the speed of light in vacuum, 
$e$ is the magnitude of the electron charge, $u = [(c^2/3)(1 + \sigma T^3/4n_0)]^{1/2}$ is the sound speed in a 
relativistically hot pair plasma, $\sigma = 4\pi^2/45\hbar^3c^3$, and $\hbar$ is the Planck constant divided by 
$2\pi$. A similar system has been previously derived for a cold plasma \cite{shukla-stenflo99}, and recently 
for a warm dust-laden plasma \cite{tsintsadze-etal06}.

A kinetic description of the radiation propagation may be obtained by the introduction of the Wigner 
density matrix (see e.g.\ \cite{wigner,moyal,loudon,mendonca}

\begin{equation}\label{eq:wigner}
  \rho_{ij}(\bm{r},t,\bm{k},\omega) = \frac{1}{(2\pi)^4}\int d\tau\,d\bm{\zeta}\,
    \exp(i\bm{k}\cdot\bm{\zeta} - i\omega\tau)
    \langle
      A_i^*(\bm{r} + \bm{\zeta}/2, t + \tau/2) A_j(\bm{r} - \bm{\zeta}/2, t - \tau/2)
    \rangle ,
\end{equation}
where the angular brackets denote the ensemble average \cite{klimontovich}. The Wigner density matrix
generalizes the the classical distribution function, and satisfies

\begin{equation}\label{eq:norm}
  I \equiv \langle|\bm{A}|^2\rangle = \int d\omega\,d\bm{k}\,\rho
\end{equation}
where $\rho = \mathrm{Tr}(\rho_{ij})$ is the trace of the state density matrix. We note that for a monochromatic pump
wave $\bm{A} = \bm{A}_0\exp(i\bm{k}_0\cdot\bm{r} - i\omega_0t) + \mathrm{c.c.}$, we have $I = 2|\bm{A}_0|^2$. 
As there is no polarization mixing and the ponderomotive force term only depends on $I$, it will here 
suffice to use $\rho$ as our radiation field description. Applying the
time derivative to the definition (\ref{eq:wigner}), using Eq.\ (\ref{eq:radiation}), and taking the trace 
of the resulting equation we obtain

\begin{equation}\label{eq:kinetic}
  \omega\partial_t\rho + c^2\bm{k}\cdot\bm{\nabla}\rho + \omega_{\rm p}^2\left(\frac{4}{3}\frac{n_1}{n_0} 
      - \frac{e^2}{m^2(T)c^4}I \right)\sin\left[ 
        \frac{1}{2}\left(
          \stackrel{\leftarrow}{\partial}_t\stackrel{\rightarrow}{\partial}_{\omega} - \bm{\nabla}\cdot\bm{\nabla}_k
        \right)
      \right]\rho = 0
\end{equation}
noting that the term proportional to the squared plasma frequency in Eq.\ (\ref{eq:radiation})
only contributes with a constant phase factor and therefore does not appear in Eq.\ (\ref{eq:kinetic}). 
The equations (\ref{eq:sound}), (\ref{eq:norm}), and (\ref{eq:kinetic}) form a closed system describing 
the nonlinear interaction of sound waves and partially coherent radiation in pair plasmas
with ultrarelativistic temperatures.

Next, we analyze the modulational instability \cite{sharma,ss} of the coupled system of equations (\ref{eq:sound}), 
(\ref{eq:norm}), and (\ref{eq:kinetic}), by letting $n_1 = \tilde{n}_1\exp(i\bm{K}\cdot\bm{r} - i\Omega t)$ and 
$\rho = \rho_0(\bm{k},\omega) + \tilde{\rho}\exp(i\bm{K}\cdot\bm{r} - i\Omega t)$, where $|\tilde{\rho}_1| 
\ll \rho_0$ and $\bm{K}$ and $\Omega$ are the perturbation wave vector and frequency, respectively. 
We then linearize with respect to the perturbations to obtain the nonlinear dispersion relation

\begin{equation}\label{eq:disprel}
  \Omega^2 - K^2u^2 = -\frac{\omega_{\rm p}^2e^2}{48T^2}\left[ 
    K^2c^2  - \frac{3}{2}(\Omega^2 - K^2u^2) 
  \right]\int d\omega\,d\bm{k}\,\frac{\rho_{0+} - \rho_{0-}}{\Omega\omega - \bm{K}\cdot\bm{k}c^2} ,
\end{equation}
where $\rho_{0\pm} \equiv \rho_0(\bm{k} \pm \bm{K}/2,\omega \pm \Omega/2)$. For a monochromatic
background distribution function $\rho_0(\bm{k},\omega) 
= I_0\delta(\bm{k} - \bm{k}_0)\delta(\omega - \omega_0)$, the expression (\ref{eq:disprel}) 
reduces to \cite{stenflo-shukla}

\begin{eqnarray}
%&&
   \Omega^2 - K^2u^2 = -\frac{I_0\omega_{\rm p}^2e^2}{24T^2} \left[ 
    K^2c^2 - \frac{3}{2}(\Omega^2 - K^2u^2)  
  \right]\bigg(
    \frac{1}{D_-}
%\nonumber \\ &&\qquad\qquad 
    + \frac{1}{D_+} 
  \bigg) ,
  \label{eq:mono}
\end{eqnarray}
where 

\begin{equation}
  D_{\pm} \equiv  K^2c^2 - \Omega^2 \mp 2\omega_0(\Omega - \bm{K}\cdot\bm{v}_g) ,
\end{equation}
and $\bm{v}_g = \bm{k}_0c^2/\omega_0$. We note that (\ref{eq:mono}) agrees with (4) in 
Ref.\ \cite{stenflo-shukla} for decay processes when $D_- \approx 0$.

Next, we consider the case of partial coherence. We will here restrict our consideration to the one-dimensional 
case. Assuming that the background electromagnetic wave has a random phase $\varphi(z)$ such that 
$\langle \exp[-i\varphi(z + \zeta/2)]\exp[i\varphi(z - \zeta/2)]\rangle = \exp(-\Delta|\zeta|)$, 
the quasi-particle distribution takes the Breit--Wigner form \cite{loudon,breit-wigner}

\begin{equation}
  \rho_0(k, \omega) = \frac{I_0}{\pi}\frac{\Delta}{(k - k_0)^2 + \Delta^2}\delta(\omega - \omega_0) .
\end{equation}
Here $\Delta$ is the half-width of the distribution centered around $k_0$. Then the dispersion 
relation (\ref{eq:disprel}) can be written as

\begin{equation}
%&&
   \Omega^2 - K^2u^2 = -\frac{I_0\omega_{\rm p}^2e^2}{24T^2} \left[ 
    K^2c^2 - \frac{3}{2}(\Omega^2 - K^2u^2)  
  \right]\bigg( 
    \frac{1}{D_- + 2iK\Delta c^2} + \frac{1}{D_+ - 2iK\Delta c^2}
%    \frac{1}{\Omega^2 - K^2c^2 - 2\omega_0[\Omega - K(v_g - i\Delta c^2/\omega_0)]}
%\nonumber \\ &&\qquad\qquad 
%    + \frac{1}{\Omega^2 - K^2c^2 + 2\omega_0[\Omega - K(v_g  - i\Delta c^2/\omega_0)]} 
  \bigg) .
\end{equation}

We next normalize to dimensionless variables according to $\omega_0 \rightarrow \omega_0/\omega_{\rm p}$, 
$\Omega \rightarrow \Omega/\omega_{\rm p}$, $K \rightarrow Kc/\omega_{\rm p}$, 
$\Delta \rightarrow \Delta c/\omega_{\rm p}$, $u \rightarrow u/c$, $v_g \rightarrow v_g/c$, 
and $I_0 \rightarrow I_0e^2/24T^2$.
In the case of the decay instability, we neglect the self-modulation of the electromagnetic wave, and 
we also have $D_- \approx 0$. Neglecting the $\Omega^2$ term in $D_-$, we have plotted the growth rate 
$\Gamma$ of the decay instability in Fig. 1. In Fig. 2 we have plotted the growth rate for the modulational 
instability. In both Figs. 1 and 2 the damping effect of the spatial spectral broadening can be seen. 
We have chosen the normalized parameter values $\omega_0 = 1$, $I_0 = 0.1$, $v_g = 0.1$, and $u = 0.2$ in 
both Figs. 1 and 2. 

%%%%%%% FIG 1 %%%%%%%
\begin{figure}
  {\includegraphics[width=.8\textwidth]{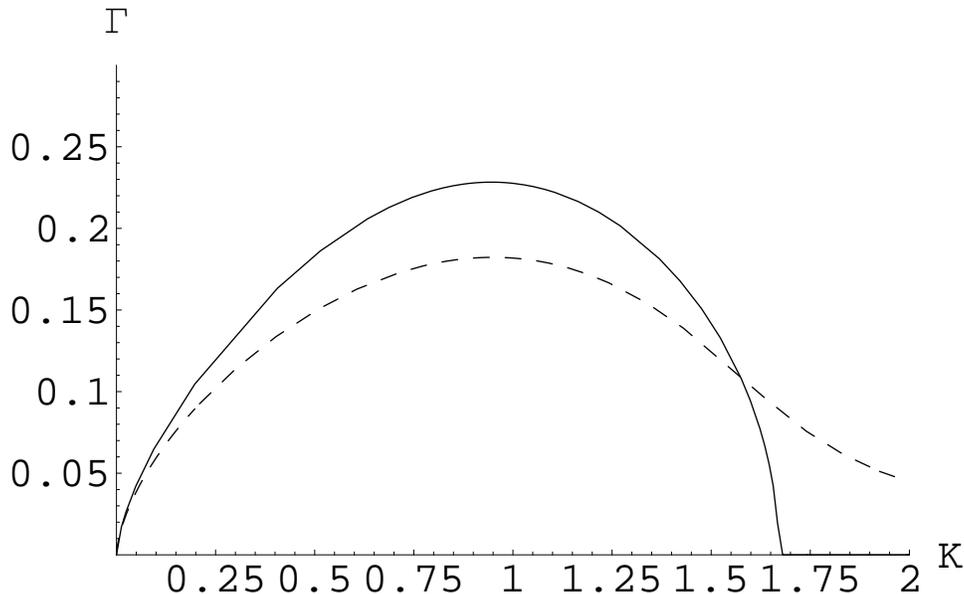}}
  \caption{The decay growth rate plotted as a function of $K$. The full curve has $\Delta = 0$, 
    while the dashed curve has $\Delta = 0.25$.}
\end{figure}
%%%%%%%%%%%%%%%%%

%%%%%%% FIG 2 %%%%%%%
\begin{figure}
  {\includegraphics[width=.8\textwidth]{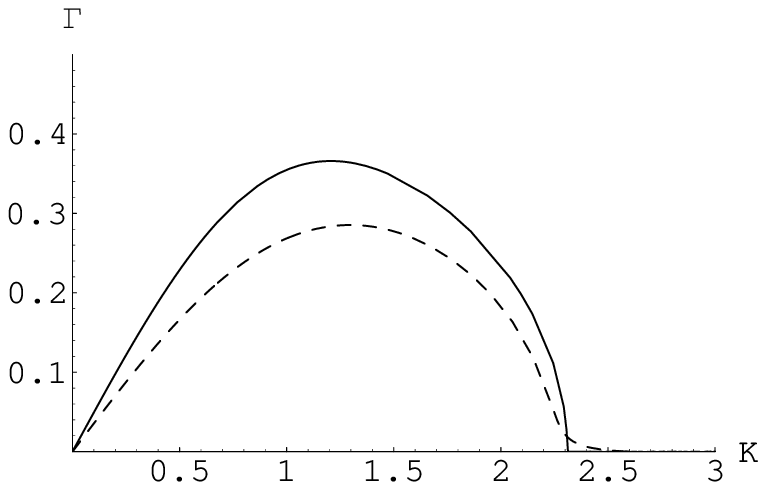}}
  \caption{The modulational instability growth rate plotted as a function of $K$. The full curve has $\Delta = 0$, 
    while the dashed curve has $\Delta = 0.25$.}
\end{figure}
%%%%%%%%%%%%%%%%%

We also plot the case of an over-dense pair plasma, such that $\omega_0 = 0.2$ in Fig. 3. 
The other parameters remain the same as above.

%%%%%%% FIG 3 %%%%%%%
\begin{figure}
  \includegraphics[width=.8\textwidth]{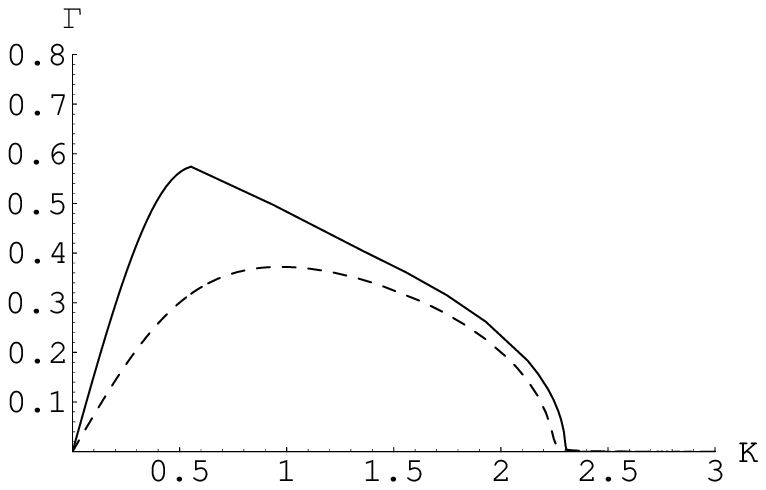}
  \caption{The modulational instability growth rate plotted as a function of $K$ in the case of an 
  over-dense pair plasma ($\omega_0 = 0.2$). The full curve has $\Delta = 0$, 
    while the dashed curve has $\Delta = 0.25$.}
\end{figure}
%%%%%%%%%%%%%%%%%

In Fig. 4 we display the modulational instability growth rate for an under-dense pair plasma, 
such that $\omega_0 = 2$. The remaining parameters remain the same as above. 

%%%%%%% FIG 4 %%%%%%%
\begin{figure}
  \includegraphics[width=.8\textwidth]{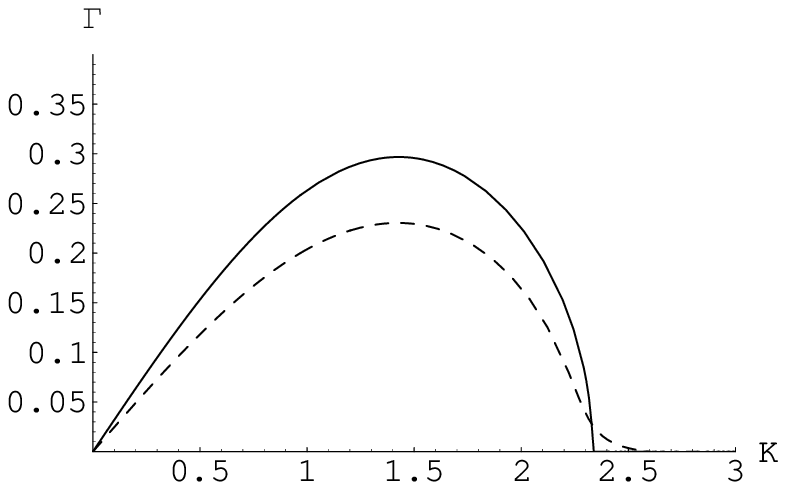}
  \caption{The modulational instability growth rate plotted as a function of $K$ in the case of 
  an under-dense pair plasma ($\omega_0 = 2$). The full curve has $\Delta = 0$, 
    while the dashed curve has $\Delta = 0.25$.}
\end{figure}
%%%%%%%%%%%%%%%%%

To summarize, we have consider the nonlinear propagation of spatial broadband electromagnetic waves 
in a relativistically hot electron-positron plasma. For this purpose, we have introduced the Wigner
transformation on a pair of equations comprising the nonlinear Schr\"odinger equation and 
a radiation pressure driven sound equation. As a result, we obtain a wave kinetic equation 
that is appropriate to derive a nonlinear dispersion relation. The latter exhibits that 
the growth rates of the decay and modulational instabilities are reduced when electromagnetic waves 
a broad spatial bandwidth. Hence, the electromagnetic waves would be able to propagate over long 
distances without losing their energy. Our results should be useful in understanding the 
nonlinear propagation of partially coherent electromagnetic waves in the early universe as well as 
in forthcoming laser-plasma interaction experiments.

This research was partially supported by the Swedish Research Council.


\begin{thebibliography}{99}

\bibitem{asseo}
E.\ Asseo, Plasma Phys.\ Control.\ Fusion \textbf{45}, 853 (2003).

\bibitem{beskin-etal}
V. S. Beskin, A. V. Gurevich, and Ya. N. Istomin, \textit{Physics of the Pulsar Magnetosphere} 
(Cambridge, 1993).

\bibitem{berezhiani}
V. I. Berezhiani, D. D. Tskhakaya, and P. K. Shukla, Phys. Rev. A {\bf 46}, 6608 (1992).

\bibitem{mourou-etal} 
G. A. Mourou, T. Tajima, and S. V. Bulanov, \rmp \ \textbf{78}, 309 (2006).

\bibitem{marklund-shukla}
M.\ Marklund and P. K. Shukla, \rmp\ \textbf{78}, 591 (2006).

  \bibitem{Piran} 
  T. Piran, Phys. Rep. {\bf 314}, 575 (1999); Rev. Mod. Phys. {\bf 76},
  1143 (2004).

\bibitem{bengt}
B. Eliasson and P. K. Shukla, Phys. Rep. {\bf 422}, 225 (2006).

\bibitem{bingham}
R. Bingham, J. T. Mendon\c{c}a, and P. K. Shukla, Plasma Phys. Control. Fusion
{\bf 46}, R1 (2004).

  \bibitem{Sturrock} 
  P. A. Sturrock, Astrophys. J. {\bf 164}, 529 (1971).

  \bibitem{Arons} 
  J. Arons and E. T. Scharlemann, Astrophys. J. {\bf 231}, 854 (1979).

  \bibitem{Michel} 
  F. C. Michel, Rev. Mod. Phys. {\bf 54}, 1 (1982).
  
  \bibitem{burke-etal}
  D. L. Burke, R. C. Field, G. Horton-Smith \textit{et al.}, Phys. Rev. Lett. \textbf{79}, 1626 (1997).
  
  \bibitem{bamber-etal}
  C. Bamber, S. J. Boege, T. Koffas \textit{et al.}, Phys. Rev. D \textbf{60}, 092004 (1999). 

  \bibitem{Campbell} 
  E. M. Campbell and W. J. Hogan, Plasma Phys. Control. Fusion {\bf 41}, B39 (1999).

  \bibitem{Cowan} 
  T. E. Cowan, M. D. Perry, M. H. Key {\it et al.}, Laser Part. Beams {\bf 17}, 773 (1999).

\bibitem{stenflo}
L. Stenflo, P. K. Shukla, and M. Y. Yu, Astrophys. Space Sci. {\bf 117}, 303 (1985).

\bibitem{shukla-etal}
P. K. Shukla, N. N. Rao, M. Y. Yu, and N. L. Tsintsadze, Phys. Rep. \textbf{135}, 1 (1986).

\bibitem{popel}
S. I. Popel, S. V. Vladimirov and P. K. Shukla, Phys. Plasmas {\bf 2}, 716 (1995).

\bibitem{shukla}
P. K. Shukla (ed.), \textit{Nonlinear Plasma Physics}, Physica Scripta \textbf{T82} (1999).

\bibitem{stenflo1}
L. Stenflo, Physics Scripta \textbf{14}, 320 (1976).

\bibitem{lominadze-etal}
J. G. Lominadze, L. Stenflo, V. N. Tsytovich \textit{et al.}, Physica Scripta \textbf{26}, 455 (1982).

\bibitem{gedalin-etal}
M. E. Gedalin, J. G. Lominadze, L. Stenflo \textit{et al.}, Astrophys. Space Sci. \textbf{108}, 393 (1985).

\bibitem{stenflo-shukla}
L. Stenflo and P. K. Shukla, Phys. Plasmas \textbf{9}, 4413 (2002).

\bibitem{shukla-stenflo99}
P. K. Shukla and L. Stenflo, Phys. Plasmas \textbf{6}, 633 (1999).

\bibitem{tsintsadze-etal06}
N. L. Tsintsadze, Z. Ehsan, H. A. Shah, and G. Murtaza, Phys. Plasmas \textbf{13}, 072103 (2006).

   \bibitem{wigner}
  E. P. Wigner, Phys. Rev. \textbf{40}, 749 (1932); 
  
  \bibitem{moyal}
  J. E. Moyal, Proc. Cambridge Philos. Soc. \textbf{45}, 99 (1949). 
  
  \bibitem{loudon}
  R. Loudon, \textit{The Quantum Theory of Light} (Oxford University Press, 2000). 
  
  \bibitem{mendonca}
  J. T. Mendon\c{c}a, \textit{Theory of Photon Acceleration} (IOP Publishing, Bristol, 2001).
  
  \bibitem{klimontovich}
  Yu. L. Klimontovich, \textit{The Statistical Theory of Non-Equilibrium Processes in a Plasma} 
  (Pergamon Press, Oxford, 1967). 
  
  \bibitem{sharma}
  R. P. Sharma and P. K. Shukla, Phys. Fluids {\bf 26}, 87 (1983).

  \bibitem{ss}
  P. K. Shukla and L. Stenflo, Phys. Fluids {\bf 28}, 1576 (1985).
  
  \bibitem{breit-wigner}
  G. Breit and E. Wigner, Phys. Rev. \textbf{49}, 519 (1936).

\end{thebibliography}
\end{document}